# On the Angular Width of Diffractive Beam in Anisotropic Media

Edwin H. Lock

(Kotel'nikov Institute of Radio Engineering and Electronics of Russian Academy of Sciences, Fryazino branch)

2-D diffraction patterns arising in the far-field region were investigated theoretically for the case, when the plane wave with non collinear group and phase velocities is incident on the wide slit in opaque screen with arbitrary orientation. This investigation was carried out by consideration as an example of magnetostatic surface wave diffraction in tangentially magnetized ferrite slab. It was deduced the universal analytical formula, which one can use to calculate the angular width of diffractive beam in any 2-D anisotropic geometries for the waves of various nature. It was shown, that in 2-D anisotropic geometries this width may be not only more or less then the value $\lambda_0/D$ ($\lambda_0$ – wavelength of incident wave, $D$ – length of slit), but it also may be equal to zero in certain conditions.

## Introduction

As it is known, different kind of waves, propagating in various media, are characterized by common physical laws. An examples of such well-known common laws for isotropic media are the laws of geometrical optics and the formula, describing angular width of diffractive beam (arising as a result of incidence of plane wave on the wide slit in opaque screen) as a ratio of incident wavelength $\lambda_0$ and slit length $D$.

For anisotropic media the laws of wave propagation, reflection, and refraction (in geometrical optics approximation) are determined by such mathematical properties of the wave isofrequency dependence (ID)[1] as the existence of asymptotes, centers of symmetry, extremum and inflection points on this dependence, the number of symmetry axes and the single- or multi-valuedness of this dependence. If ID of the wave possesses by some of these properties, then propagation, reflection, and refraction of waves can be accompanied by such phenomena as the impossibility of wave propagation in some directions or within some angular sector, nonreciprocality of propagation (when counter-propagating rays have different parameters), unidirectional propagation (when for a given ray there exists no counter-propagating ray and, in some cases, there is only a single direction, in the vicinity of which the energy can be transferred), negative reflection and refraction (when the incident, reflected, and refracted rays are on the same side of the normal to the interface), the emergence of two (or more) reflected or refracted rays, the absence of reflection altogether, and the irreversibility of ray paths in reflection or refraction (see [1] and references, reported there).

It is evidently to suppose, that mathematical properties of the wave ID also determine the angular width of diffractive beam in anisotropic media and structures. So, the question is appears: is it possible to deduce some universal formula, describing angular width of diffractive beam in anisotropic media 2-D geometries (similar to the formula for isotropic media)? How it is seen from analysis of scientific literature, diffractive phenomena in anisotropic media were investigated mainly for electromagnetic wave in plasma, for light in optical crystals, for acoustic waves and for dipole spin waves, named usually magnetostatic waves (MSWs) [2]. However, it seems that such formula is absent for all these waves at the moment (probably, because of mathematical difficulties). So an attempt is taken to obtain such universal formula through the study of MSW diffraction in ferrite slab.

---

[1] ID is also known as "section of wavevector surface", "section of the isoenergy surface" and "equifrequency line" [1].

**Statement of the Problem and Main Results.**

Let's consider an infinite ferrite slab of thickness $s$ magnetized to saturation by a tangent homogeneous magnetic field $\mathbf{H_0}$. The ferrite slab is characterized by well-known magnetic permeability tensor, and ambient half-spaces have magnetic permeability equaling to 1. To describe MSW in ferrite slab let's use magnetostatics equations rot $\mathbf{h}$ = 0 and div $\mathbf{b}$ = 0 and introduce magnetostatic potential $\Psi$ in agree with formula $\mathbf{h}$ = grad$\Psi$ [2]. Then consider diffraction pattern, arising in the most general case – when plane surface MSW with frequency $f_0$ and non collinear wave vector $\mathbf{k_0}$ and group velocity vector $\mathbf{V_0}$ is incident on the slit of length $D$ in opaque thin screen with arbitrary orientation (Fig.1). For convenient study of this problem there are used Cartesian $\Sigma_C = \{x; y; z\}$ and Polar $\Sigma_P = \{x; r; \varphi\}$ coordinate systems, coupled with ferrite slab anisotropic directions (with direction of external magnetic field $\mathbf{H_0}$), and another Cartesian and Polar systems, coupled with orientation of the slit and screen ($\Sigma'_D = \{x; y'; z'\}$ and $\Sigma'_P = \{x; r; \varphi'\}$ respectively).

Due to potential $\Psi$ is scalar function, the study of MSW diffraction becomes significantly simpler and we can follow in general by the widely known analytical ways, used for isotropic media [3 - 5]. However, it is necessary to take into account anisotropic character of MSW propagation in ferrite slab. In particular, the next important mentions must be made.

1. Let's suppose, that the result of plane MSW incidence onto slit is equal to appearance of distribution of secondary elementary MSW sources with various phases along slit line, and to estimate the resulting action of these sources in the far-field region we must integrate contributions from all these elementary sources (i.e. calculate superposition of all these sources).

2. Action of every secondary source is defined by magnetic potential excited by this source, and, consequently, to calculate resulting action of all sources in the far-field region it is possible to summarize potentials excited by all sources (whereas every potential is scalar function).

3. As a distinct from the solving of similar problems in isotropic media, we always will to deal with two directions $\varphi$ and $\psi$ (or with $\varphi'$ и $\psi'$ directions in $\Sigma'_D$ and $\Sigma'_P$ coordinate systems) corresponding to orientation of the wave vector $\mathbf{k}$ and group velocity vector $\mathbf{V}$ respectively. While conditions, determining constructive interference of the secondary sources, are formulated for their wave vectors (i.e. for $\varphi$ or $\varphi'$ directions), the MSW energy is transferred (when this constructive interference is realized) along the corresponding group velocity vector direction (i.e. along $\psi$ or $\psi'$ directions). So for mathematical description of studied diffractive problem it is necessary to use two points at infinity - $P_{\mathbf{k}}$ and $P_{\mathbf{V}}$: the direction from the slit centre to the $P_{\mathbf{k}}$ point is coincided with some orientation $\varphi$ of secondary sources wave vectors, and the direction from the slit centre to the $P_{\mathbf{V}}$ point – with corresponding group velocity vector orientation $\psi$ (Fig.1). Thus, ID of MSW $k(\varphi)$ in Polar coordinate system (see inset on Fig.1) and corresponding dependence $\psi(\varphi)$ are important for our further consideration. The sake of simplicity let's assume, that dependencies $k(\varphi)$ and $\psi(\varphi)$ are single-valued functions (for surface MSW in free ferrite slab this assumption is always correct), and, therefore there is only one point $P_{\mathbf{V}}$, corresponding to every point $P_{\mathbf{k}}$. Let's assume also, that inverse dependence $\varphi(\psi)$ is single-valued function too.

With these assumptions for the case $D/\lambda_0 >> 1$ after some analytical calculations it is possible to find that angular distribution of total (summary) magnetic potential from all secondary MSW sources is determined by expression of ~ sin$\Phi/\Phi$ (similar to analogous expression for isotropic media), but the phase function $\Phi$ is more complex, than in isotropic media

$$\Phi(\varphi(\psi)) = \pi \frac{D}{\lambda_0}\left[\sin(\varphi_0 - \theta) - \frac{k(\varphi(\psi))}{k_0}\sin(\varphi(\psi) - \theta)\right], \qquad (1)$$

where $k(\varphi)$ and $\varphi(\psi)$ – dependences, characterized certain wave in anisotropic medium or structure (for considered example this wave is surface MSW in ferrite slab), $\varphi_0$ and $k_0 = 2\pi/\lambda_0$ – parameters of incident plane MSW, $D$ and $\theta$ – length and orientation of the slit (see Fig.1). Since sin$\Phi/\Phi$ is rapidly oscillating function we may neglect by the slowly changing factor (similar to Kirchhoff factor) in beam width calculations. Finding difference between directions $\psi$, corresponding to the cases $\Phi = 0$ ($\sin\Phi/\Phi = 1$) and $\Phi = \pi$ ($\sin\Phi/\Phi = 0$), it is possible to deduce formula, describing *absolute* angular width of main diffractive beam $\Delta\psi$ (at level 0.5). However in anisotropic media it is more useful *relative* angular width of main diffractive beam $\sigma$, which is related with $\Delta\psi$ by the simple formula $\sigma = \Delta\psi/(\lambda_0/D)$. Relative width $\sigma$ show, how much absolute angular beam width $\Delta\psi$ differs from the similar width $\lambda_0/D$ in isotropic media, if slits lengths and lengths of incident waves are the same. Relative angular width $\sigma$ of main diffractive beam is described by the formula

$$\sigma = \left| \frac{\frac{d\psi}{d\varphi}(\varphi_0)}{\frac{\frac{dk}{d\varphi}(\varphi_0)}{k_0}\sin(\varphi_0 - \theta) + \cos(\varphi_0 - \theta)} \right| = \left| \frac{\frac{d\psi}{d\varphi}(\varphi_0)}{\frac{\frac{dk}{d\varphi}(\varphi_0)}{k_0}\sin\varphi'_0 + \cos\varphi'_0} \right|, \qquad (2)$$

where $d\psi/d\varphi$ and $dk/d\varphi$ – derivatives values of functions $\psi(\varphi)$ and $k(\varphi)$ respectively at $\varphi = \varphi_0$. There is used sign of modulus, because it is convenient to characterize angular width by positive numbers, like a distance. If the wave vector $\mathbf{k_0}$ of incident MSW is perpendicular to the slit line ($\varphi'_0 = 0$) denominator in (2) is equal to unity and formula (2) looks most simple. In this case for isotropic media (whose ID is circumference, dependence $\psi(\varphi)$ has the form $\psi = \varphi$ and $d\psi/d\varphi \equiv 1$) we get $\sigma = 1$ and arrive to well-known expression $\Delta\psi = \lambda_0/D$.

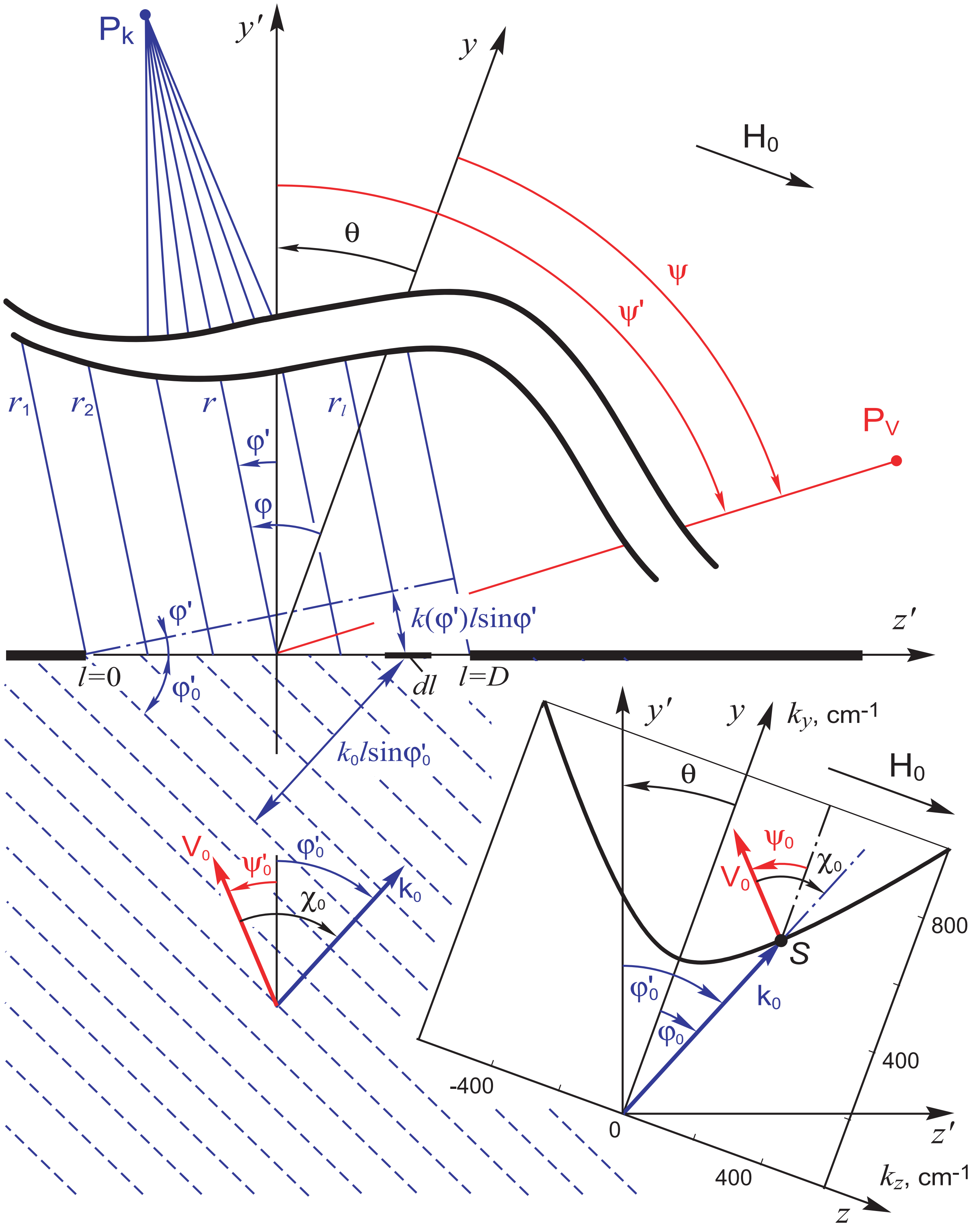

Fig.1. Geometry of the problem in the plane of ferrite slab. Phase fronts of plane incident MSW are shown by dashed lines. Inset: half of ID for surface MSW with $f_0$ = 2900 MHz in free ferrite slab (point S corresponds to incident MSW with non collinear vectors $\mathbf{k_0}$ and $\mathbf{V_0}$).

In spite of deduction method there is a hope, that formula (2), deduced for MSW, will be valid for any anisotropic media and structures for 2-D geometries, including metamaterial structures (whose IDs is differ from circumference too). As it is seen from the formula (2) an unusual phenomenon may be appear in anisotropic 2-D geometries: if incident wave is characterized by such value $\varphi_0$ that $d\psi/d\varphi = 0$ at $\varphi = \varphi_0$, then $\sigma = \Delta\psi = 0$! It means, that diffractive beam conserve its wide during propagation! Mention must be made, that not any wave has ID with point(s), where $d\psi/d\varphi = 0$, but such points present on ID for surface MSW in free ferrite film. Calculations, based on formula (2) together with numerical computations (when the difference between directions $\psi$, corresponding to $\Phi = 0$ and $\Phi = \pi$ are calculated on agree with (1)) are shown in the Fig. 2 for the case $\psi'_0 = 0$, that corresponds to all geometries, where incident MSW vector $\mathbf{V_0}$ is normal to the slit line and orientation of incident MSW wave vector $\mathbf{k_0}$ is changed (i.e. angle $\chi_0$ is varied from -90$^o$ to +90$^o$). MSW's ID for $f_0$ = 2330 MHz has the points, where $d\psi/d\varphi = 0$ (at $\chi_0 = \pm$ 73$^o$) and, as it is seen from Fig. 2, if incident MSW is characterized by angle $\chi_0 = \pm$ 73$^o$, then relative angular beam width $\sigma = 0$. Mention must be made also, that $\sigma \to 0$ at $\chi_0 \to \pm$ 90$^o$ due to $dk/d\varphi \to \infty$ in (2) near ID asymptote.

**Summary**

2-D problem of diffraction on the slit for MSW with non collinear group and phase velocities is solved. It is shown, that angular width of diffractive beam is determined by both parameters of incident wave and mathematic properties of wave's ID. It is deduced the universal analytical formula, which one can use to calculate the angular width of diffractive beam in any 2-D anisotropic geometries for the waves of various nature (including waves in metamaterials). It is shown, that in anisotropic media (including metamaterials) the angular width of diffractive beam may be equal to zero at the certain conditions (when the incident wave is characterized by such point on the wave's ID, where $d\psi/d\varphi = 0$).

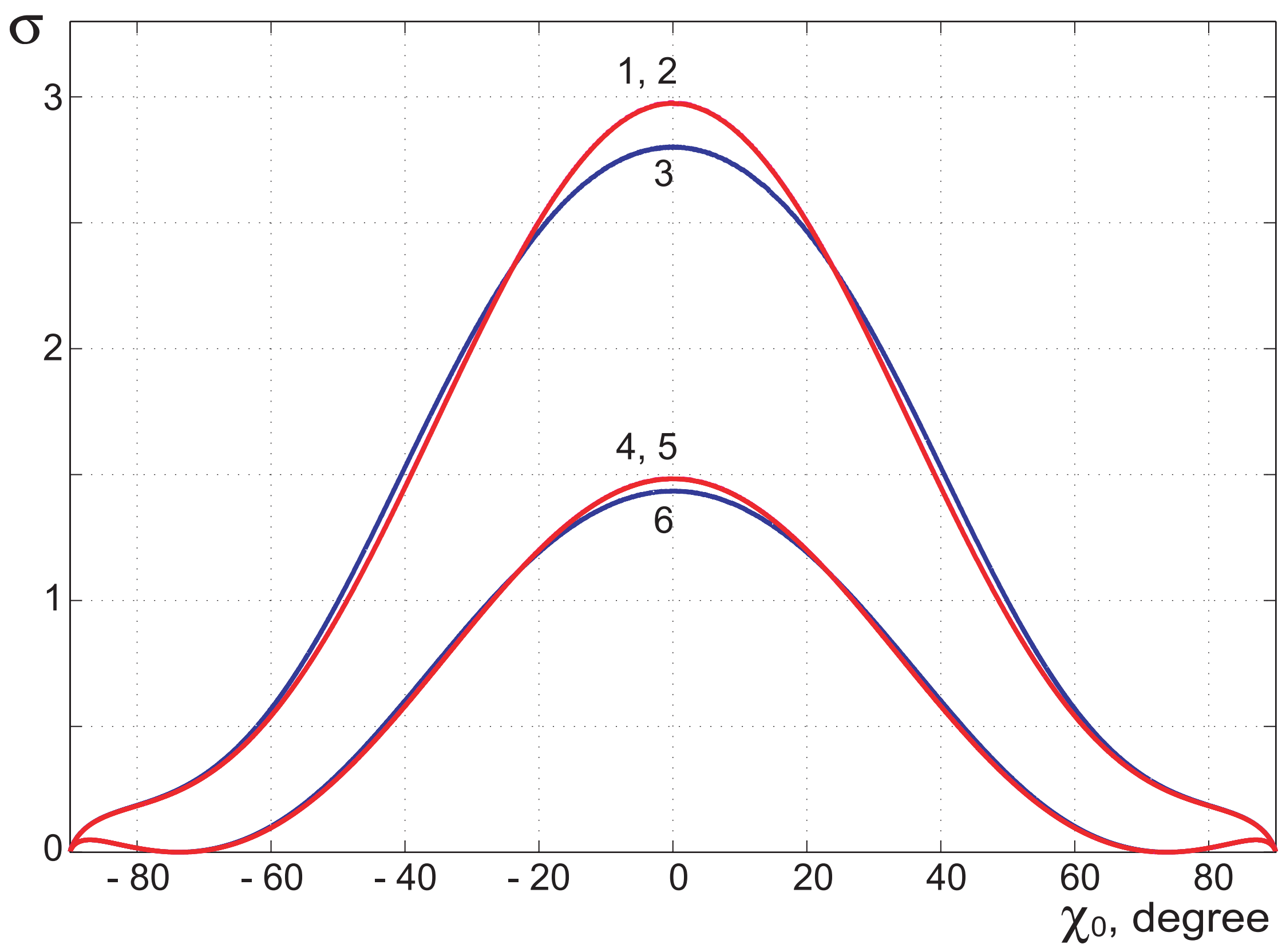

Fig.2. Relative angular width $\sigma$ of transmitted diffractive beam versus angle $\chi_0$ between vectors $\mathbf{V_0}$ and $\mathbf{k_0}$ of incident MSW for $\psi'_0 = 0$ and $H_0 = 300$ Oe, saturation magnetization $4\pi M_0 = 1750$ Gs, $s = 10$ μm: *1 – 3* – for $f_0 = 2900$ MHz, *4 – 6* – for $f_0 = 2330$ MHz. *1* and *4* – calculations, using formula (2); *2*, *5* and *3*, *6* – numerical computations for $\lambda_0/D = 0.01$ and $\lambda_0/D = 0.1$ respectively.

This work is partially supported by the Program "Development of the Scientific Potential of High School" (project No. 2.1.1/1081).